\documentclass[conference]{IEEEtran}
\usepackage{cite}
\usepackage{amsmath,amssymb,amsfonts}
\usepackage{algorithmic}
\usepackage{graphicx}
\usepackage{textcomp}
\usepackage{xcolor}
\usepackage{fancyhdr}
\usepackage[hyphens]{url}
\usepackage{soul}
\usepackage{siunitx}
\usepackage{booktabs,colortbl}
\usepackage{array,multirow,makecell}
\usepackage{pgfplotstable}
\usepackage{caption}
\usepackage{tikz}
\usepackage{xspace}

\pgfplotsset{compat=1.8}

\definecolor{rulecolor}{RGB}{0,71,171}
\definecolor{tableheadcolor}{gray}{0.92}

\pgfplotsset{compat=1.8}

\definecolor{rulecolor}{RGB}{0,71,171}
\definecolor{tableheadcolor}{gray}{0.92}

\newcolumntype{R}[1]{>{\raggedleft\arraybackslash }b{#1}}
\newcolumntype{L}[1]{>{\raggedright\arraybackslash }b{#1}}
\newcolumntype{C}[1]{>{\centering\arraybackslash }b{#1}}

\def\BibTeX{{\rm B\kern-.05em{\sc i\kern-.025em b}\kern-.08em
    T\kern-.1667em\lower.7ex\hbox{E}\kern-.125emX}}

\newcolumntype{P}[1]{>{\hspace{0pt}}p{#1}}

\fancypagestyle{firstpage}{
  \fancyhf{}

  \fancyfoot[C]{\thepage}
}

\graphicspath{{Figs/}}
\newcommand{\catalog}{LocMap\xspace}

\definecolor{mygreen}{RGB}{0,90,0}

\newcommand{\fig}[1]{Figure~\ref{#1}\xspace}

\newcommand{\sect}[1]{Section~\ref{#1}\xspace}

\newcommand{\ballnumber}[1]{\tikz[baseline=(myanchor.base)] \node[circle,fill=.,inner sep=1pt] (myanchor) {\color{-.}\bfseries\footnotesize #1};}

\title{Reducing Load Latency with \\Cache Level Prediction} 
\author{Majid Jalili, and Mattan Erez\\ The University of Texas at Austin \\
\{majid,mattan.erez\}@utexas.edu}

\begin{document}

\maketitle
\thispagestyle{firstpage}
\pagestyle{plain}


\begin{abstract}
High load latency that results from deep cache hierarchies and relatively slow main memory is an important limiter of 
single-thread performance. Data prefetch helps reduce this latency by fetching data up the hierarchy before it is requested by load instructions. However, data prefetching has shown to be imperfect in many situations. We propose cache-level prediction to
complement prefetchers. Our method predicts which memory hierarchy level a load will access allowing the memory loads to start 
earlier, and thereby saves many cycles. The predictor provides high prediction accuracy at the cost of just one cycle added 
latency to L1 misses. Experimental results show speedup of 7.8\% on generic, graph, and HPC applications over a baseline 
with aggressive prefetchers. 
\end{abstract}

\section{Introduction}
\label{sec:intro}
Low memory-load latency is critical for high-performance computing applications. Achieving low load latency is challenging 
because latency has been trending up as cache hierarchies grow in capacity and complexity. Recent Intel processors, for 
example, have estimated second- and third-level cache (L2 and L3) latencies of 12 and 40 cycles, respectively \cite{InteSpec}. 
The levels of the hierarchy are typically looked up in sequence, starting from the first-level cache (L1) and proceeding 
through second- and third-level caches. If the data is not found in any cache, it is fetched from memory. Deep cache 
hierarchies generally improve performance, but can result in higher load latencies when caches do not successfully filter 
requests, only adding lookup delays \cite{MultiLevelCache,FilterCache, ACCORD}.

Prefetchers somewhat mitigate the latency impact of level-by-level lookup by moving data between levels prior to the execution 
of load instructions~\cite{BOP,STeMS,ISB,DCPT,SBOOE,IMP,SPPV2,PIF,SlimAMPM,AMPM}. Although prefetchers improve system performance, many loads are still exposed to sequential-lookup delays. Previous work has demonstrated miss coverage of just 24\% and 40\% for SPEC CPU 2006 \cite{MetaPrefetcher} and CloudSuite \cite{Bingo,DominoPrefetcher}, respectively. 

A naive approach to reduce cache hierarchy latency is to look up caches in parallel. However, this increases energy consumption and requires over-provisioning tag array ports and possibly on-chip bandwidth. Alternatively, the entire hierarchy, tag store, 
TLBs, and coherence protocol can be completely redesigned to allow Direct-to-Data, lookup-free cache access \cite{D2D}. Unlike 
such costly and radical solutions, we propose a \emph{memory hierarchy level predictor} that enables directly looking up the 
cache level where a block resides \emph{with minimal changes to the memory hierarchy}.

The level predictor gives us the best of both worlds: the reduced access latency of non-sequential (parallel) lookup while maintaining the low access costs and simplicity of a sequential hierarchy. In this way, \emph{level prediction} (LP) improves performance for those loads that inevitably miss in the cache despite advanced prefetchers and replacement policies. Our analysis of a large set of benchmarks running on an Intel Skylake processor demonstrates that many applications, including from the graph analytics and scientific computing domains are likely to benefit from non-sequential cache access, even when cache hit rates are high.
 
%

On an L1 miss, the level predictor (within each core) predicts which memory levels to target, bypassing some levels and 
occasionally indicating partial parallel lookups across memory levels. The bypass and parallel access hardware mechanisms reuse 
structures that already exist in current sequential lookup implementations, requiring only small modifications to control 
logic. Any mispredictions that incorrectly bypass a level 
or that perform unnecessary parallel 
accesses are also handled with cache controller modifications and reuse existing structures: unnecessary parallel accesses are 
terminated by modifying existing address matching-logic on the return path, and incorrect bypasses are rectified by the 
cache-coherence directory reissuing requests to the correct level.



We propose a novel per-core level predictor comprised of two components: (1) a metadata cache of a global \emph{location map} (\emph{\catalog})
that holds  possibly-stale information about the level of each  memory block, and (2) a compact history-based \emph{Popular
Level Detector} predictor that is used on metadata cache misses. The \catalog requires 2 bits per 512-bit cache block to
indicate its memory level. It is organized as a flat table in system-reserved physical memory. It is accessed with the same
granularity as the data cache---information for 256 cache blocks is fetched on every \catalog access after a metadata cache
miss. We find that just a 2 KiB metadata cache of the \catalog provides a high hit rate while keeping access energy and energy low.

The compact Popular-Level Detector generates a level prediction when level information is not available in the \catalog metadata cache (a metadata miss). This is required because the long delay of fetching \catalog information would otherwise render the level prediction useless. The history predictor requires just a 32-bit per-level register for counting the number of hits to each level and simple logic that implements a heuristic of combining the counters to generate a level prediction.

Level prediction is distinct from prior work on cache miss prediction (also referred to as memory access prediction)~\cite{MattanHitMiss,MissMap,MoinL4,Mostly-Clean,PeirHitMiss}. While at a high level LP is analogous to simultaneous miss prediction at all cache levels (excluding L1, which has a very short latency), a practical level predictor \emph{must be architected from scratch}. We find that combining or stacking miss predictors squanders opportunities to utilize cross-level information, necessitating a new predictor design. Notably, prior work did not consider the impact of advanced prefetchers. When combined with better prefetchers, prior miss predictors exhibit poor accuracy and much larger resource requirements. Given the stringent area and power constraints necessary for the gains of level prediction to outweigh its costs, again, a new predictor architecture is required.


To summarize our main contributions:
\begin{itemize}
\item 
  We demonstrate that many graph 
  analytics and scientific applications benchmarks can benefit from non-sequential lookup; including applications with a high 
  cache hit ratio.
  
\item 
  We architect and evaluate an effective, yet low-cost level predictor that operates in each core on the L1 cache miss path
  and substantially outperforms cache miss predictors when incorportaing prefetchers. 
  
  \item We incorporate level prediction keeping with minimal microarchitectural changes; level misrepdictions are handled by utilizing the cache coherence directory and existing address-matching logic.

  \item We evaluate our proposed method and compare it to an ideal baseline and two state-of-the-art \cite{D2D, MissMap}. Overall, level prediction improves the performance of both general, scientific, and graph analytics applications by 7.8\% on average over a baseline with aggressive prefetchers. Level prediction also reduces the cache hierarchy energy consumption by 18\% on average, as level prediction eliminates many unnecessary miss lookups.

\end{itemize}


\section{Motivation and Background}
\label{sec:motivation}
Modern processors employ three or more levels of cache. The cache levels are accessed in sequence. The first level cache (L1, the highest level cache) is looked up first, and in the case of a miss the request is forwarded to L2, and then L3 (the lowest level cache). If the data is not found in any cache, it is read from main memory. To provide non-blocking access, each cache has miss status holding registers (MSHRs). MSHRs track outstanding misses, allowing the processor to continue execution while a request is serviced. MSHRs also  coalesce different targets to the same cache block.

Current processors use a distributed directory to maintain coherence. The directory tracks the locations of any block in all on-chip caches to redirect requests and force write-backs and invalidations when necessary.
The directory plays a significant role in our design and we rely on it to resolve level mispredictions. Note that directory lookup does not replace cache level prediction. First it is located in or near the LLC tags, and thus far from the L2 cache. Second, the cost of the directory lookup is high and thus not beneficial from both latency and energy perspectives.

Many applications are memory-latency bound. To study the effectiveness of the level-by-level lookup strategy, we compare the number of misses at different levels.  If the application shows good locality or has access patterns that are detected by prefetchers, the number of demand load misses decreases significantly from one level to the next. Our analysis below demonstrates that while sequential level lookup indeed works well for many applications that exhibit this type of behavior, other applications suffer from unnecessary lookups as either L2 does not successfully filter requests, or L3 provides no substantial additional benefit over the L1 and L2 caches.

\begin{figure}[tb]
\centering
\includegraphics[width=\columnwidth]{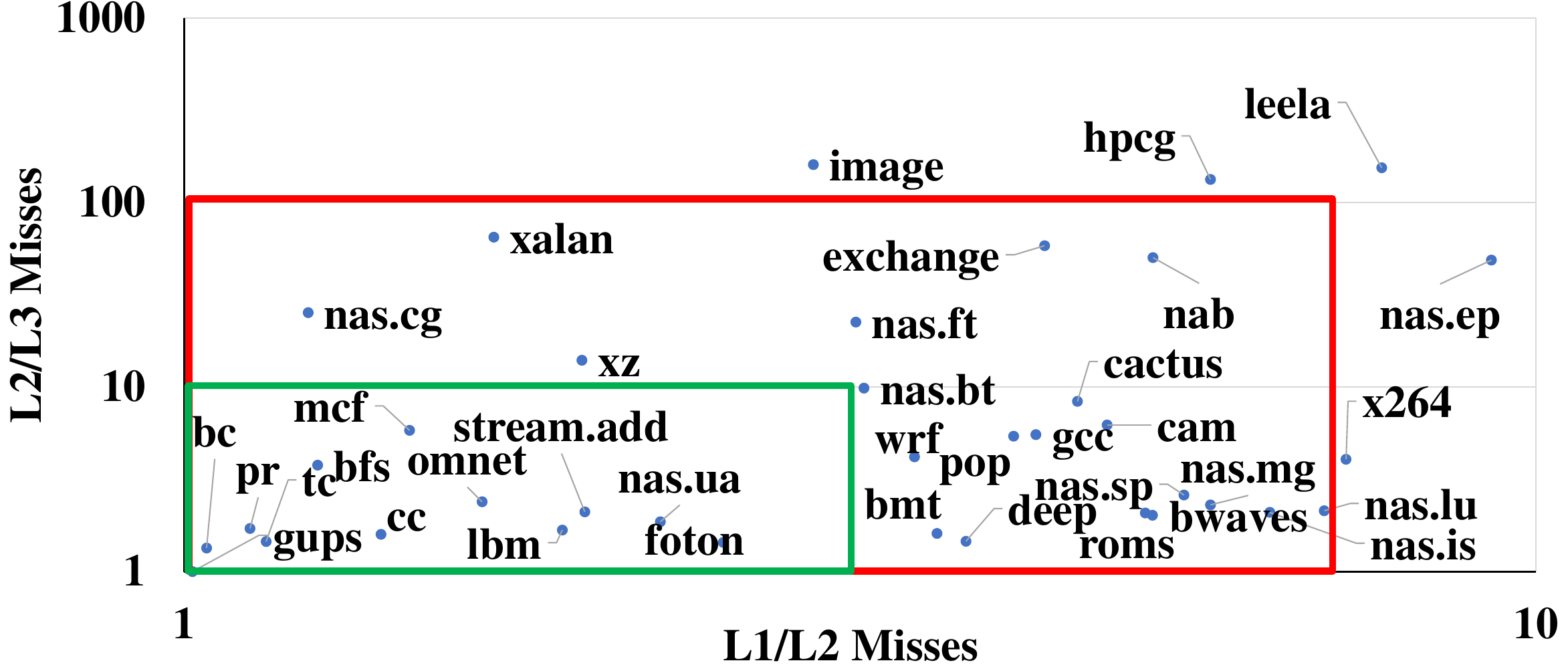}
  \caption{The x- and y-axes on this log-log plot represent the effectiveness of L2 and L3 at filtering misses, respectively. Each application is plotted according to the ratio of misses filtered by each level compared to the level above it. Applications further to the top and right work best with sequential access, while those toward the bottom left stand to benefit most from non-sequential level-predicted lookups.}
\label{fig:appsDomain}
\end{figure}

The high-level insights from our analysis are summarized in Figure~\ref{fig:appsDomain}. The figure plots each evaluated application in terms of its L2 and L3 effectiveness---the filtering capability of each level of cache. The x-axis (log scale) is the ratio of L1 to L2 misses and points further on the right indicate applications for which L2 more effectively filters L1 misses from reaching L3, thus indicating that looking up L2 before L3 is the right strategy. Similarly, the y-axis (log scale) represents the effectiveness of L3 and higher points are for applications where misses from L2 are mostly hits in L3. Data on cache effectiveness is collected on a 3.2GHz Intel Core i7-8700 CPU for SPEC CPU 2017 and NAS Parallel Benchmarks applications, the GAPBS graph analytics benchmark suite, and for the \emph{hpcg}, \emph{gups}, \emph{stream}, \emph{spmv}, and \emph{bmt} kernels (see Section~\ref{sec:eval} for more details on methodology).

Using this figure, we roughly classify applications into three categories: (1) applications outside the red box are a good fit for sequential level lookup and are unlikely to benefit from level prediction, (2) applications inside the green box, for which non-sequential lookup is likely to offer significant latency reductions, and (3) applications between the boxes where we expect that L2, L3, or levels can be occasionally bypassed for modest performance gains. We observe that not only graph analytics applications exhibit poor cache effectiveness, but that many other applications are likely to benefit from level prediction. We provide a more detailed analysis below.

\subsection{Sequential Lookup Effectiveness}
\begin{figure*}[b!t]
\centering
\includegraphics[width=2.35in]{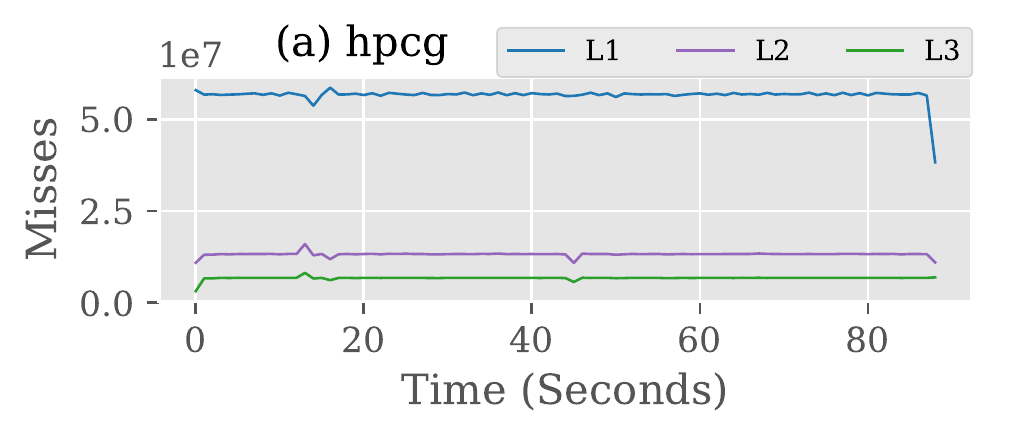}
\includegraphics[width=2.35in]{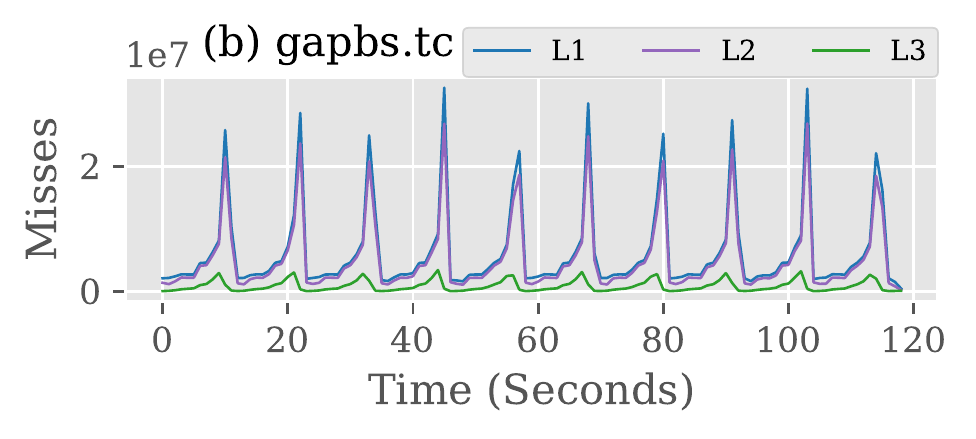}
\includegraphics[width=2.35in]{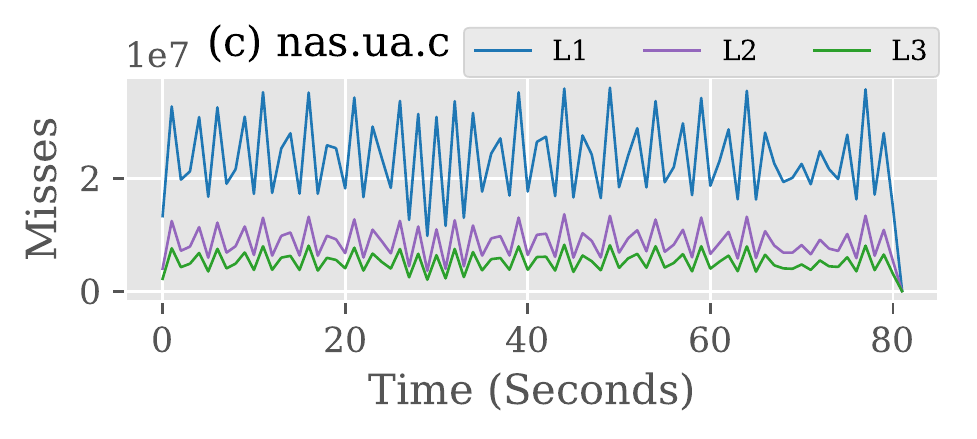}

\includegraphics[width=2.35in]{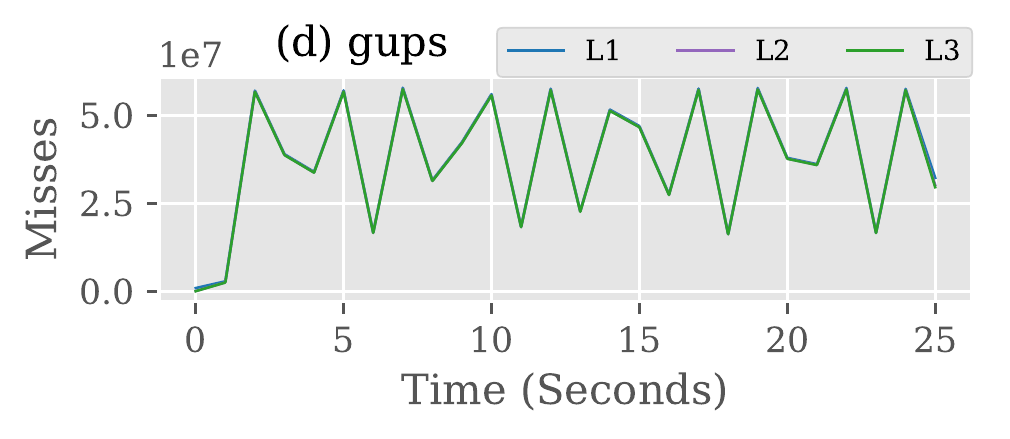}
\includegraphics[width=2.35in]{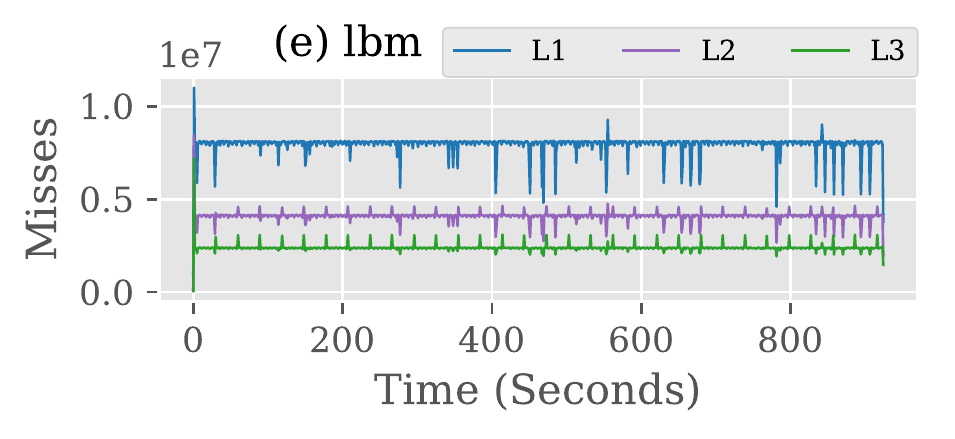}
\includegraphics[width=2.35in]{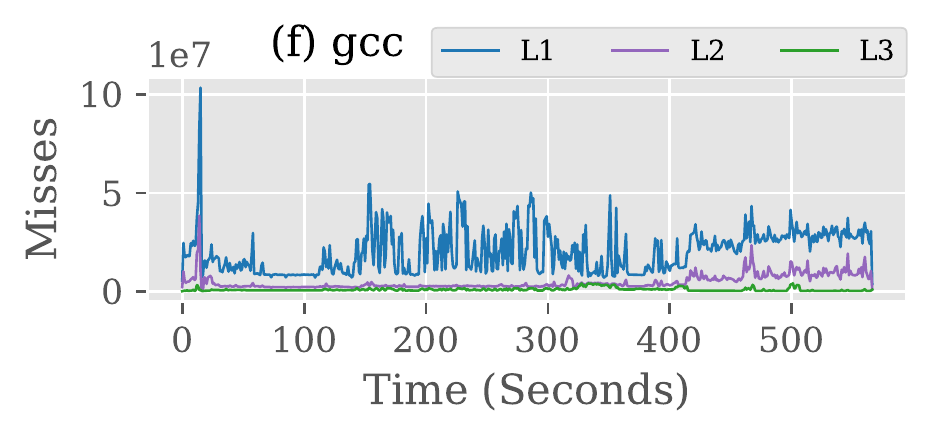}
  \caption{Miss trace of several applications across their execution. A gap between the miss rate (number of misses per time window) of different cache levels indicates effective miss filtering, while lines that are close to one another suggest a cache level lookup can be bypassed.}
\label{fig:motivationl}
\end{figure*}

Figure~\ref{fig:motivationl} provides more detail on the number of misses at each cache level across the execution duration of several applications. We expect sequential level lookup to be the best design choice when caches reduce the number of misses significantly. Figure~\ref{fig:motivationl}(a) exemplifies such behavior and shows that for \emph{hpcg} (which falls outside the red box of Figure~\ref{fig:appsDomain}), the number of misses significantly decreases after L2 (3$\times$ reduction) and also after L3 (a further 2$\times$ decrease). This behavior is consistent throughout execution, meaning serial lookup performs well over the course of execution. 

Many other applications, however, suffer from serial lookup. Figures~\ref{fig:motivationl}(b-f) show applications where either L2, L3, or both are not effective. If the miss rate at any level is very high, the access latency increases unnecessarily by looking up that level. This is especially a problem when an intermediate level, has a very high miss rate. This behavior is common in graph applications, which frequently exhibit poor hit rates at L2 and moderate hit rates at L3~\cite{Sam_IMP}. For example, for GAPBS Triangle Count (\emph{tc} in Figure~\ref{fig:motivationl}(b)), we observe a similar rate of L2 and L1 misses, indicating that L2 is ineffective. L3 only moderately reduces the number of misses. Therefore, almost all L2 accesses and the majority of L3 lookups are redundant and only increase memory access latency.  

A case where L3 cache is ineffective is shown in Figure~\ref{fig:motivationl}(c) where the number of misses at L3 is roughly the same as at L2, despite the fact that L3 is 48$\times$ larger.
Considering the fact that the L3 is large and has a high access latency, level-by-level lookup only squanders CPU cycles by looking up L3 for every single access. This problem can be exacerbated when the application has a random access pattern as in \emph{gups} (Figure~\ref{fig:motivationl}(d)). Generally, random behavior impairs both prefetchers and caches, and thus almost all references to caches waste cycles that could otherwise be spent directly looking up main memory.  

While it intuitively seems that sequential accesses should be the best choice for applications with simple access patterns, that is not necessarily the case. For example, Figure~\ref{fig:motivationl}(e) shows that despite both L2 and L3 reducing the misses for an easy-to-prefetch application like \emph{619.lbm}, the number of misses at each level is still high. Meaning a substantial fraction of the memory requests still needs to traverse the memory hierarchy level-by-level wasting many cycles. 

Application behavior may change during execution, and thus using a static strategy to lookup the caches is not optimal.
Figure~\ref{fig:motivationl}(f) shows the miss rates for \emph{602.gcc} from SPEC CPU 2017. L2 is not very effective at the early stages of execution (0--25 seconds), is beneficial in filtering out requests from 20--400 seconds, and decreases to lower effectiveness for the rest of execution. The hardware level predictor can exploit this phase-dependent behavior and skip looking up levels of the hierarchy (L2 in this case) when they do not provide benefit.

\begin{figure}[t]

\centering
\includegraphics[width=2.8in]{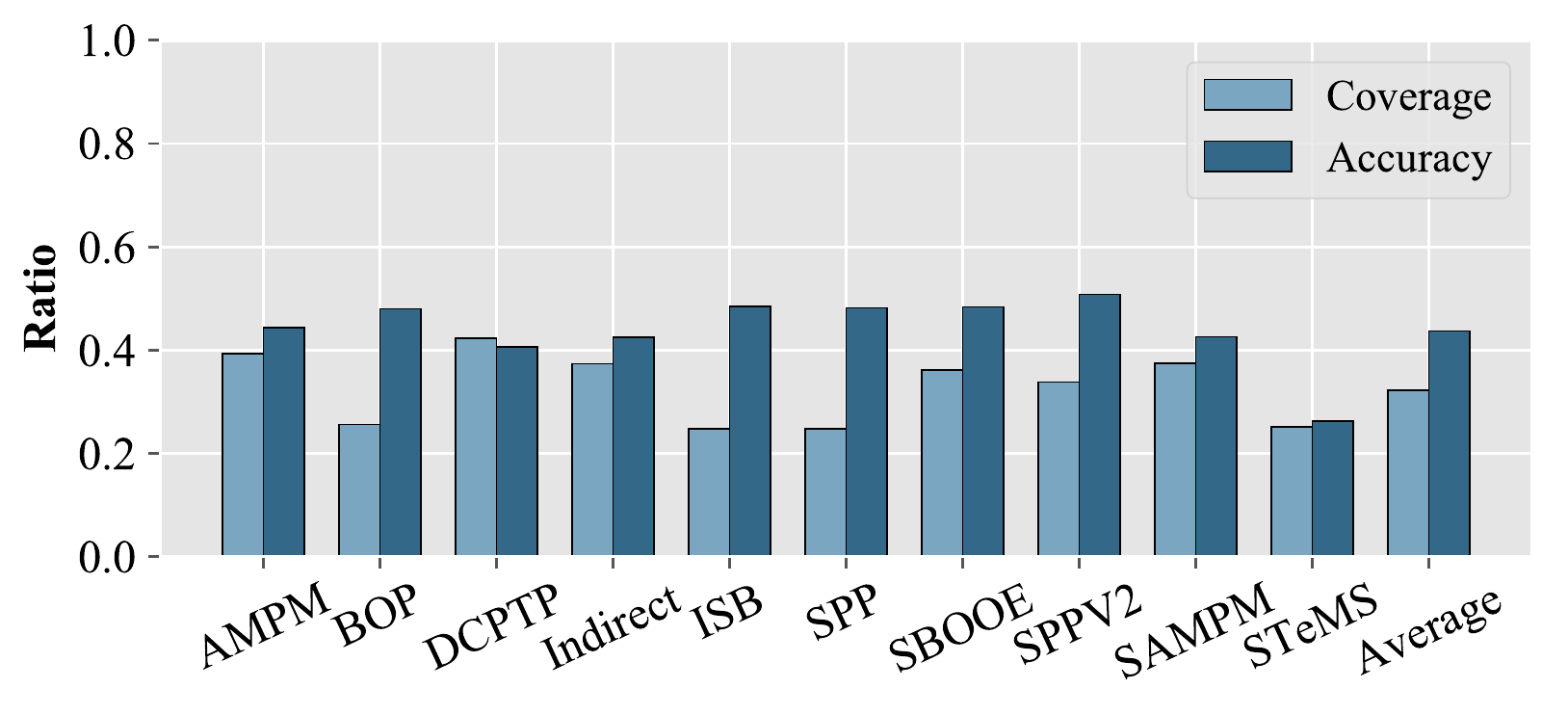}
\vspace{-0.5em}
 \caption{Coverage and accuracy of LLC prefetchers. Coverage is the fraction of misses eliminated by prefetching, and accuracy is the fraction of useful prefetches. In the best cases (e.g., DCPT \mbox{\cite{DCPT}}), 40\% of misses are eliminated but with high overhead of 40\% inaccuracy. High inaccuracy of prefetchers makes level prediction very challenging.}
\label{fig:Coverage_all}
\end{figure}

\noindent\textbf{Prefetchers.}
Despite progress in designing prefetchers, many misses are left uncovered. Figure~\ref{fig:Coverage_all} shows the simulated coverage and accuracy of  numerous state-of-the-art academic prefetchers \cite{BOP,STeMS,ISB,DCPT,SBOOE,IMP,SPPV2,PIF,SlimAMPM,AMPM}. As a matter of fact, in the best-case scenarios, prefetchers can eliminate up to 50\% of LLC misses, meaning many accesses still access slow main memory. The average accuracy is also low (50\%-60\%), meaning many unnecessary blocks are fetched and evicted with a possible negative impact on performance. We offer a new mechanism that complements advanced prefetchers and replacement policies, rather than replacing them. We attempt to handle those misses that even state-of-the-art prefetchers leave for main memory.

\section{Level Prediction Microarchitecture}
\label{sec:proposed}
\subsection{Predictor Design Considerations}

\noindent\textbf{Levels to Predict.}
Without loss of generality, assume that there are four memory hierarchy levels, and any given block can reside in L1, L2, L3, and main memory. Out of these, we exclude L1 as a prediction target for two reasons. First, many processors use virtually-indexed 
physically-tagged (VIPT) L1 caches (L1 and TLB are accessed concurrently), meaning the physical address needed for skipping L1 is not available during the L1 lookup itself. 
Second, the L1 is tightly integrated with the core pipeline, and we choose not to disrupt its timing or design.
%
We do include L2 as a level prediction target because our evaluation results indicate that skipping L2 lookups offers substantial speedup in many of the applications that benefit from level prediction. Additionally, our level predictor 
is simple and scalable, and thereby can be simply extended to predict more levels, if required.

\noindent\textbf{Resource and power constraints.}
A level predictor is integrated with each and is queried on every L1 miss. As a result, it must be area-efficient to be realistically implemented and operate with very low power to provide benefits even when skipping the fairly low-power L2 lookup operation. Our analysis shows that larger predictors increase energy consumption higher than the baseline, even if the prediction accuracy is high. Hence, we target predictor structures that are small and lower power.

\noindent\textbf{Level Prediction Approach.}
\label{sec:miss_predictors}
A level predictor must determine whether to skip L2 lookup, L3 lookup, or both. While, this is distinct from prior work on miss predictors, our LP architecture is inspired by such work. We group miss predictors into three categories: PC + miss history~\cite{MattanHitMiss,MoinL4}, address + miss history~\cite{Mostly-Clean}, and a ``miss map'' approach~\cite{MissMap, PeirHitMiss}.


The miss-history predictors rely on classic binary predictors (e.g., TAGE~\cite{TAGE}) to predict the level of a cache block based on its observed hit/miss history. However, LP is conceptually more straightforward because the location of a cache block is available at any time within the tag arrays. The miss-map approach, therefore, uses a table~\cite{MissMap} or a Bloom filter~\cite{PeirHitMiss} to track which blocks are in a specific cache.
 
One approach for LP is to ``stack'' separate L2 and L3 miss predictors. However, after extensive experimentation, we conclude that this stacking approach performs poorly with both lower accuracy and high power consumption than our proposed level predictor (detailed analysis in Section \ref{sec:eval}). Instead, it is beneficial to extend miss prediction to true level prediction by combining multi-level information in a single structure and tracking per-level history and level location, rather than single-level history and presence.

Specifically, we enhance the TAGE-based address+history miss predictor of Sim et al.~\cite{Mostly-Clean}. We replace each TAGE counter entry with three counters in each entry to keep track of each level (L2, L3, and main 
memory). On prediction, if an entry is found in a TAGE table, we use a 
Popular Levels Detector heuristic that uses the three counters to predict a cache level; we describe this heuristic in greater detail later in this section. If an entry is not found in any TAGE table, we follow a level-by-level traversal.

Our design-space exploration identifies several additional crucial issues with all prior designs, even when extended to LP. \emph{We find that the MissMap approach requires tables that are too large for the tight resource budget of the LP while the history-based predictors work poorly in the presence of advanced prefetchers, unless significant additional storage resources are made available.}


\noindent\textbf{Impact of Prefecthers.}
  One important reason for the poor performance of stacked miss predictors is the presence of advanced prefetchers. While prior publications report very high prediction accuracy for a range of history-based PC-indexed~\cite{MattanHitMiss,MoinL4} and address-indexed~\cite{Mostly-Clean,PeirHitMiss} miss predictors, prior evaluations did not include sophisticated prefetchers. We find that prefetchers add ``noise'' to the hit/miss history and substantially degrade history-based predictor accuracy. We evaluate a range of predictors with and without advanced prefetchers and find that prefetchers necessitate much larger prediction structures.

Furthermore, there is an opportunity for coordinating the prefetcher and level predictor: the prefetcher may update the level predictor about data movement within the hierarchy to improve prediction accuracy. We add such updates to both our LP-extensions of prior miss predictors and to our own LP design described below. We find that even with such updates, history-based predictors perform poorly because either the additional histories from prefetches or the additional entries introduced by prefetch updates overwhelm the prediction tables at sizes that are reasonable for the tight LP resource constraints.

\subsection{Cache Level Predictor Microarchitecture}
Given the level prediction considerations described above, we opt for a novel LP microarchitecture that extends the MissMap~\cite{MissMap} approach in three critical ways (\fig{fig:design}). First, we extend the MissMap to a \emph{location map} (\catalog), which provides the location of each block (L2, L3, or MEM), requiring 2 bits of metadata per block. Second, the original MissMap is implemented as a cache and loses all information about a block when a MissMap entry is evicted. We find that this either requires a very large MissMap table or leads to low LP accuracy. Instead, we implement the \catalog as an in-memory table containing location information for every block in physical memory, which is cached in a small metadata cache. The long-term location information is then available even after an eviction. Third, because the metadata cache is small and full \catalog access has high latency, we add a small history-based Popular Levels Detector that provides fast level prediction on a metadata cache miss. This history predictor requires just three counters.


The level predictor is 
attached to the L2 bus and communicates with the L2 and L3 caches, which report fill and dirty eviction events to the level
predictor. These events are used to update the \catalog. In addition, the three counters of the Popular Levels Detector receive hit and miss 
signals from the caches. We describe the components and their operation below.


\subsection{\catalog and Location Tracking}


The \catalog is a flat table in system-reserved physical memory and is accessed with the same granularity as the data cache. The \catalog holds the level information of all memory blocks using 2 bits of metadata (there are 3 possible levels to predict: L2, LLC and main memory). Each 64-bit \catalog entry holds location information of 32 blocks. To provide fast access to \catalog, hot metadata is cached on-chip. For 64B cache blocks, this metadata scheme incurs only {$\frac{2}{512}=0.39\%$ overhead.

\noindent\textbf{\catalog Access.} The \catalog is accessed (through the metadata cache) on every L1 cache miss, and when it is updated. Each block in physical memory is mapped to an entry in \catalog. Hence, to access the \catalog, we need to generate an address from the block's physical address. To do so, we employ a simple one-to-one mapping. We assume that the base address to the \catalog table is set by the operating system and that the memory access granularity and cache blocks are 64B. Each 64B cache block requires 2 bits in the \catalog such that information for 256 cache blocks fits into a single 64B block of the \catalog (matching the memory access and cache granularity). Hence, the memory address corresponding to a \catalog entry is$\mathit{\catalog\ Address}=\mathit{Base\ Address}+\mathit{Physical\ Address}>>14$. This physical address of a \catalog block is first looked up in the per-core \catalog metadata cache, which is filled on a miss through the data cache hierarchy and main memory. 

\noindent\textbf{\catalog Update.} Level prediction does not need to be 100\% accurate. Hence, we can carefully trade accuracy for power-efficiency. This can be achieved by updating the catalog on certain events. We update the catalog only on demand cache fills, dirty evictions, and prefetch fills that are metadata cache hits. Thus, the \catalog may hold possibly-stale information because it is not updated on all events happening in the cache. Prefetch fills that are metadata cache misses do not update the \catalog because the traffic this would incur with our aggressive prefetchers is substantial and not worth improving the already-high prediction accuracy (see Section~\ref{sec:results}). Because of the aggressive prefetchers, there are frequent clean evictions and these do not update the \catalog as well. Finally, to avoid changes to the coherence protocol and actions, coherence-induced level changes (i.e., invalidations) are also ignored. Again, staleness is tolerable because the predictor already performs well and because misprediction recovery is inexpensive.


\noindent\textbf{Metadata Cache.}
The sizing of the metadata cache is important. If the miss ratio is too high, the problem is two-fold. First, many off-chip requests are issued to update the \catalog. Second, the prediction accuracy may degrade as we have to rely on the statistical Popular Level Predictor, which is less accurate than the \catalog. At the same time, the size of the \catalog metadata cache is constrained to maintain low access latency and energy---the benefits of level prediction can easily be overwhelmed by an expensive predictor. We show this in the evaluation by demonstrating the detrimental impact of a higher-energy predictor, for example.  

In order to find a good size for the metadata cache, we measure the average energy for various capacities. We simulate a 4-way 64KB set associative L1 cache along with a 2-way metadata cache. We sweep over metadata cache sizes to study the average energy consumption.
Figure \ref{fig:metaMissRate} shows the normalized average energy for sizes of 1KB, 2KB, 4KB, and 8KB for 4 benchmark suites: SPEC CPU 2017, GAPBS, NAS, and other applications (\emph{bmt}, \emph{hpcg}, \emph{spmv}, \emph{gups}, and \emph{stream}; annotated as \emph{Others}). We observe that a 2KB metadata cache provides a reasonable tradeoff between the extra overhead and the coverage provided by the cache.

\begin{figure}[t]
\centering
\includegraphics[width=3in]{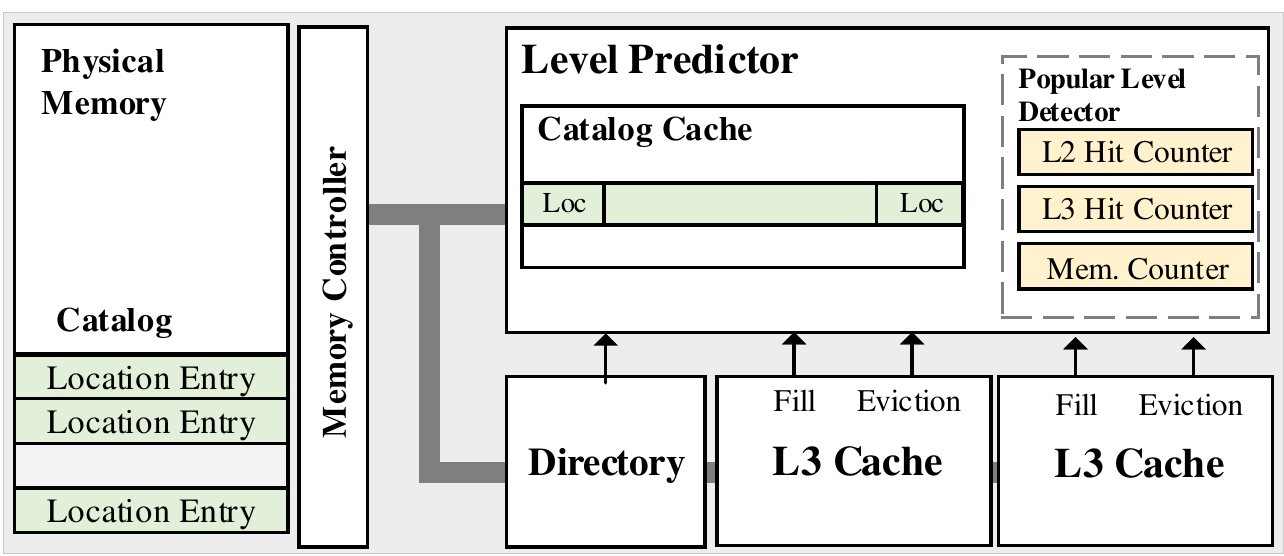}
  \caption{The proposed architecture.}
\label{fig:design}
\end{figure}

\begin{figure}[h]
\centering
\includegraphics[width=2.8in]{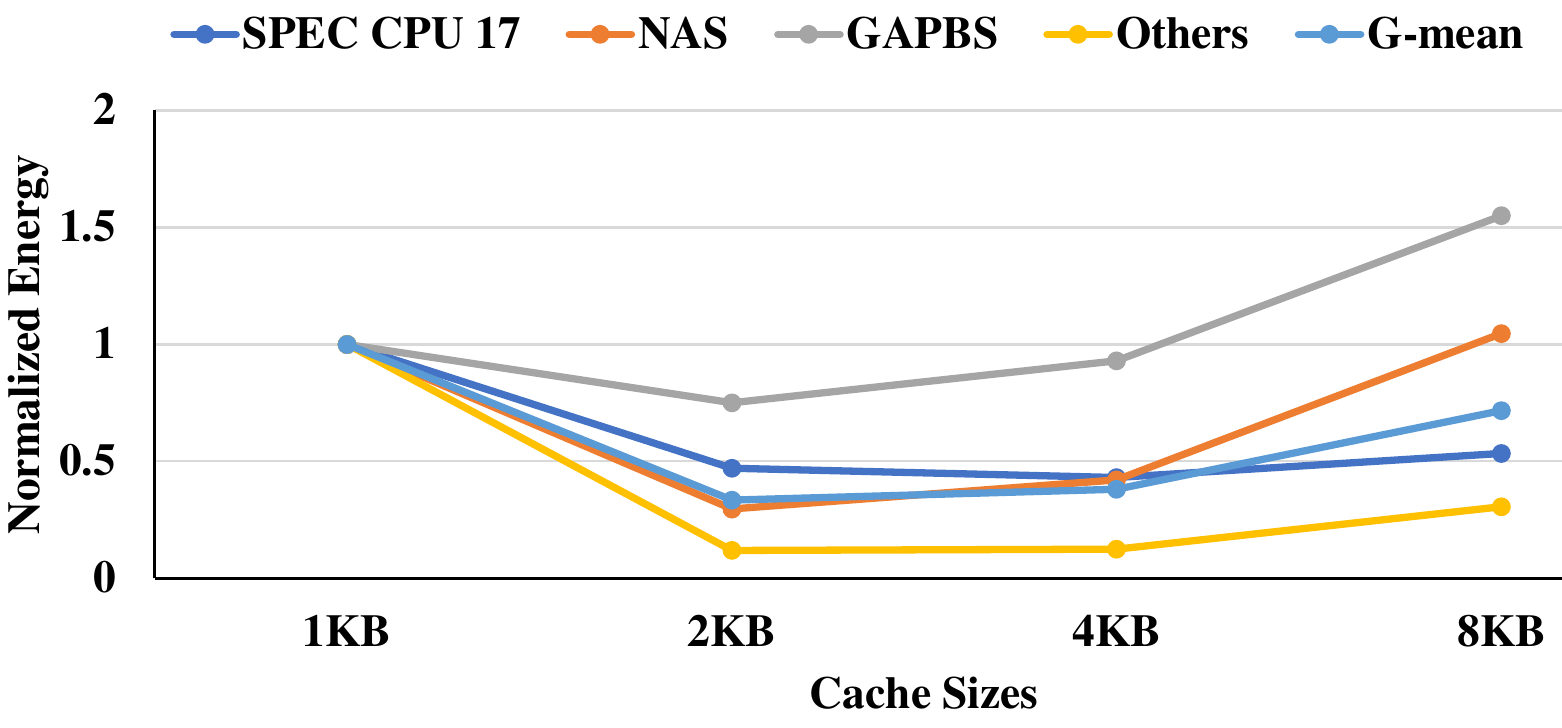}
  \caption{Normalized average energy for different cache sizes. Numbers are normalized to 1KB cache.}
\label{fig:metaMissRate}
\end{figure}
\begin{figure*}[t]
\centering
\includegraphics[width=\textwidth]{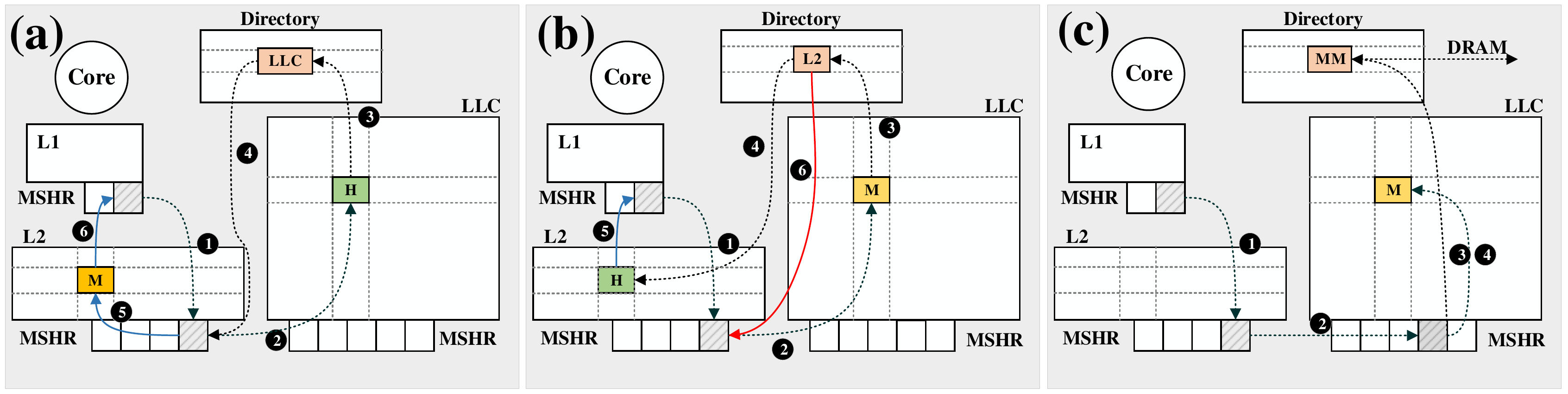}
  \caption{Example of parallel access with cache level prediction: (a) One-way correct prediction (prediction=LLC, actual location=LLC); (b) One-way wrong prediction (prediction=LLC, actual location=L2); (c) Multi-way with a wrong prediction (prediction=LLC and main memory, actual location=main memory)}.
\label{fig:access_example}
\end{figure*}

\subsection{Popular Levels Detector}

When there is a miss in the metadata cache, it is not possible to wait to fetch the \catalog entry from main memory because this takes longer than the cache lookup itself. One option is to follow a serial access pattern and predict L2 as the location of the block. However, this option is too conservative and does not cover applications with high metadata cache miss ratios such as \emph{pr}, \emph{bc}, \emph{tc} of the GAPBS benchmark suite. Hence, in conjunction with the \catalog, we devise a simple counter-based mechanism to find the most frequently accessed levels and suggest those as the prediction target(s).

Because the metadata hit rate is relatively high (95\% on average across 37 applications), we can be more aggressive on (relatively rare) misses and predict more than one level to increase the prediction accuracy. This is particularly helpful if the counters are not strongly biased toward one level. If one level is suggested, we call the prediction a single-way prediction, and a multi-way prediction otherwise. Multi-way prediction for uncertain cases increases the prediction accuracy, but requires a more complicated lookup.

We use 3 counters, one per cache level and main memory. Upon a hit at a level the corresponding counter is incremented by 1 and others are decremented by 1. This helps to rapidly find popular levels and prevents counter saturation. When a prediction is required, candidates are selected as follows. The counters are sorted and the topmost is picked as the first candidate. If its counter is higher than a threshold, then only this level is selected as the level to look up. Otherwise, the level with the second highest counter is also considered to be a possible destination (two parallel lookup targets). If again the sum of the first and the second counters does not reach a predefined threshold, the third level is also included (three parallel lookup targets). Hence, depending on the counter values the locality predictor may issue single- or multi-way predictions.

\subsection{Misprediction Detection and Recovery}
Mispredicting a level that is closer to the core than the actual cache level does not require detection and recovery, and is simply a lost opportunity to reduce latency. For example, if the block is only in main memory and LLC is suggested, then not looking up L2 saves cycles, but looking up the LLC wastes CPU cycles; MSHR entries are still allocated along the request path and will be eventually filled as the response arrives.  Therefore, this type of misprediction is safe as it does not violate correctness or functionality. 

However, mispredicting a miss and bypassing a level that has more up to date data does require recovery as stale data is fetched. For example, if data is present in L2, but only L3 is suggested by the level predictor, then stale data may be incorrectly fetched from L3. To cope with this problem, we slightly modify the directory controller. Normally, the directory is checked to make sure that private caches of other processors are not holding the data block. When level prediction bypasses L2, we also check the directory, as we may have skipped the private cache. Fortunately, this can be done effectively for free. Recent processors collocate the directory with LLC tags, meaning when the cache is looked up, the information of the directory is also available \cite{Skylakedirectory}. This enables simple misprediction detection.

Another misprediction type is where main memory is predicted while the block is actually cached. This misprediction is also detected by the directory, which is queried in any case before accessing main memory. 

We change the cache controller to raise a signal when a misprediction is detected. Misprediction recovery first issues a packet to the actual level, requesting to satisfy the pending request and then deallocates all MSHR entries past the actual level. This can be implemented as a new transaction over the shared bus.

We present examples of detection and recovery schemes from the simplest (one-way correct prediction) to the most complex (multi-way wrong prediction) case below.

\noindent\textbf{One-way correct prediction.} Figure \ref{fig:access_example}(a) shows an example for one-way correct prediction. Assume the predicted location is LLC, and the block is present in the LLC (green box). When an L1 miss occurs, an MSHR entry is allocated in L1 and the request is sent to L2 (\ballnumber{1}). Then without accessing L2, an MSHR entry (denoting an L2 miss) is allocated in L2. Hence, L2 is bypassed and the request is forwarded to the LLC (\ballnumber{2}). Note that allocating an entry in the MSHRs for bypassed levels (in this case L2) is necessary, otherwise it would be impossible to fulfill the request later on the fill path. At the LLC, the tag-store is checked (\ballnumber{3}). Since, the block already exists in the cache and the directory confirms that the block is not stale, the LLC responds to the request (\ballnumber{4}). The block is sent to L2 and the L2 MSHR is deallocated, and it is filled into L2 (\ballnumber{5}). Finally, the block is forwarded to L1 which responds to the CPU request (\ballnumber{6}). 

\noindent\textbf{One-way wrong prediction.} Figure \ref{fig:access_example}(b) shows an example where a block is present in L2, but the predictor suggests LLC. In this example, steps \ballnumber{1}, \ballnumber{2}, and \ballnumber{3} are the same as before; MSHR entries are allocated in L1 and L2 and the request is subsequently sent to the LLC. However, when the tag store is looked up in the LLC, the extended way information indicates that the block is present in L2 \ballnumber{4}. Thus, the cache controller sends a new request to L2 to fulfill the request \ballnumber{5}. In L2, the block is found and is forwarded to L1 \ballnumber{6}. Also, a signal is raised to deallocate the L2 MSHR entry \ballnumber{7}. The only modification here is to add the ability to cache controller to deallocate the MSHR entry.

\noindent\textbf{Multi-way wrong prediction.} Figure \ref{fig:access_example}(b) shows an example where LLC and DRAM have been suggested as targets by the predictor, but the block is present in DRAM. Steps \ballnumber{1}, \ballnumber{2}, and \ballnumber{3} are the same as in both the previous examples. However, when the packet reaches to LLC, both LLC and directory are accessed, and after finding the location of the block which is the main memory, the request is forwarded to the main memory. Finally, when the request comes back, it can follow the normal path that any miss sourced in DRAM can take. Note that in our design the directory and LLC tags are collocated, thus as soon as the tag is accessed, the request is sent to main memory.


\section{Evaluation Methodology}
\label{sec:eval}
\subsection{Simulated System}
We use the gem5 full-system cycle-level simulator to conduct the experiments \cite{gem5}. We model a 3-level cache hierarchy 
where L1 and L2 are inclusive and private and L3 is non-inclusive and shared. L1 and L2 are parallel caches where tag and 
data stores are accessed in parallel with access latencies of 4 and 8 cycles, respectively. L3 is a sequential cache with 
latencies of 20 cycles and 55 cycles for tag and tag+data, respectively. There is a first-level TLB of 64 entries, a second-level TLB with 3072 entries that are equally partitioned between 4KB and 2MB pages (1536 each). The L2TLB is 4-way set associative with a  4-cycles access latency. There are 2 page walkers per core.

Our goal is to evaluate level prediction with an advanced prefetch scheme. We therefore experimented extensively with a wide range of state-of-the-art prefetchers and their combinations. The highest-performing scheme overall in our experiments uses the DCPT prefetcher~\cite{DCPT} with degree 2 in L3 and tagged next-line prefetchers of degrees 1 and 2 for L1 and L2, respectively. DCPT exhibits the highest coverage and high accuracy (Figure \ref{fig:Coverage_all}) and worked well in combination with the L1 and L2 prefetchers. While these next-line prefetchers may appear simplistic, they work remarkably well given the tighter timing and accuracy needs of the smaller cache levels.


We also find that always enabling these prefetchers significantly degrades system performance for some applications (e.g., 605.mcf) because the prefetchers contend too strongly with demand requests. We therefore implement two prefetch throttling mechanisms. 
In the first scheme, we reserve 25\% of MSHR entries for demand accesses, which decreases the prefetch rate and maintains some minimum demand request service. The second throttling mechanism is that  we monitor the performance of the prefetcher periodically and disable a prefetcher when its accuracy drops below 40\%. Specifically, in each epoch of 10 million accesses, the prefetchers operate for the first 1 million accesses, then the prefetcher accuracy determines if the prefetcher remains enabled for the following 9 million accesses.

We simulate out-of-order cores with a fetch width of 4 instructions, 192 ROB entries, and   32-entry store and  32-entry load queues. The frequency of the system is set to 4GHz. 
We use a single DDR4-2400 x64 channel (one command and address bus), with timings based on a DDR4-2400 8 Gbit datasheet (Micron MT40A1G8) in an $8\times 8$ configuration. Total channel capacity is 16GB. This maintains a reasonable core-to-memory ratio for the simulations.

\subsection{Benchmarks}
We evaluate the applications of: (1) SPEC CPU 2017 \cite{spec17}, (2) GAPBS \cite{gapbs} (\emph{pr}, \emph{tc}, \emph{cc},
 \emph{bfs}, and \emph{bc}), (3) NAS (\emph{cg}, \emph{ft}, \emph{is}, \emph{mg}, and \emph{ua}) and (4) \emph{bmt},
 \emph{hpcg}, \emph{stream-copy}, and \emph{gups}. In the evaluation section we report averages for the full benchmark suites, 
 but choose to highlight 21 applications to maintain readability of figures. We pick 12 applications that we expect to highly-
 benefit from level prediction (within the green box of Figure {\ref{fig:appsDomain}}) and 9 applications that we expect to 
 exhibit smaller benefits (from within the red box).

All SPEC CPU applications are run with the reference inputs. We use the Twitter \cite{Twitter} dataset for GAPBS, with the 
exception of \emph{tc} that uses a synthetic graph of $2^{25}$ nodes. For NAS, input class \emph{C} is used. For \emph{gups}, 
we replace the random generator with the C++ built-in random generator to ensure that the table is randomly accessed. The table 
size is 8GB  and 4 million locations are accessed. We compile all benchmarks with gcc/gfortran and -O3 flags.

We use the SimPoint methodology \cite{simpoint} to find representative regions of each application. We use 2 SimPoints of 250 million instructions each and 250 million instructions of warmup. For kernels (\emph{gups}, \emph{stream}, \emph{bmt}), we annotate the code with gem5 pragmas to simulate just the region of interest.

For multi-core evaluation we use a set of multi-program and multi-threaded applications listed in Table \ref{Table:multiprogram}. We use level prediction accuracy from single core simulation to observe how different applications with 
high, medium, and low prediction accuracy interact. From application mixes, mix1, mix3, and mix5 have 2 high expected benefit 
applications (green box) and 2 expected medium benefit applications (red box), mix2 has 1 high-benefit application and 3 
medium-benefit applications (red box), and mix4 has 4 expected medium-benefit applications. For multi threads, we focus on 
GAPBS.pr with 2 and 4 threads. Given the GAPBS.pr has one the lowest single-core hit prediction accuracy, we can observe how
 level prediction accuracy changes as the contention increases, and how the accuracy is impacted as the \catalog does not update 
 the prediction table on snoop invalidations.

\begin{table}[t]
\footnotesize
\renewcommand{\arraystretch}{1}
\caption{Evaluated System Configuration.}
\label{Table:System}
\centering
\resizebox{\linewidth}{!}{%
\begin{tabular}{|p{0.6in}||p{2.3in}|}
\hline
Processor & Single and Quad-core, 4.0GHz, Ubuntu 16.04 OS. ROB:192, LQ:64, SQ:64, Fetch-width=4\\
\hline
L1 Cache & 32kB, Split I and D cache; 32KB private; 4-way; 64B line size; LRU; write-back; 1 port; 4 cycles. Tagged next line prefetcher with degree=1\\
\hline
L1 and L2 Coherency  & MOESI directory; L1 and L2 are inclusive, L3 is non-inclusive\\
\hline
L2 Cache & 256KB; 8-way; 64B line size; LRU; write-back; 2 ports; 12 cycles, tagged Next Line prefetcher with degree=2 \\
\hline
L3 Cache & 2MB single-core and 8MB multi-core; 16-way; 64B line size; LRU; write-back; 4 ports; Sequential cache (20+35). DCPT prefetcher degree of 2\\
\hline
Main Memory & 16~GB:  DDR4-2400 x64, Micron MT40A1G8 in an 8x8 configuration\\
\hline
\end{tabular} }
\end{table}

\begin{table}[t]
\renewcommand{\arraystretch}{1}
\center
\caption{Multi-program and multi-threaded applications.}
\label{Table:multiprogram}
\begin{tabular}{l }
\hline
mix1~~~~~~~GAPBS.bfs, SPEC.619.lbm, NAS.lu, bmt       \\ 
mix2~~~~~~~SPEC.654.roms, NAS.mg, SPEC.649.fotonik3d, SPEC.602.gcc \\ 
mix3~~~~~~~SPEC.620.omnetpp, GAPBS.pr, SPEC.627.cam, NAS.cg       \\ 
mix4~~~~~~~SPEC.627.cam, NAS.cg, SPEC.621.wrf, NAS.bt             \\ 
mix5~~~~~~~GAPBS.bfs, SPEC.619.lbm, SPEC.621.wrf, NAS.bt      \\ \hline
MT1~~~~~~~~GAPBS.pr with 2 threads \\
MT2~~~~~~~~GAPBS.pr with 4 threads \\
\hline
\end{tabular}
\end{table}

\subsection{System Comparisons}
To evaluate the level prediction, we compare LP to 4 systems: (1) 2KB TAGE (energy competitor) \cite{Mostly-Clean}, (2) 8KB 
TAGE (prediction accuracy competitor) \cite{Mostly-Clean}, (3) Direct-to-Data (high implementation cost, but energy and 
accuracy competitor) \cite{D2D}, and (4) Ideal system where misses do not incur any performance penalty. In Ideal, we use 0 
cycle for the miss latency with other functionalities of the simulator remaining the same. For example, bus and MSHR latencies 
are still counted for misses.  For D2D, we assume that
the \emph{Hub} is an 8-way 4KB cache. Additionally, we assume that the eTLB requires 10\% higher energy per access as it 
increases the length entries\cite{D2D}.


\section{Evaluation Results}
\label{sec:results}

\begin{figure*}[t]
\centering
\includegraphics[width=\textwidth]{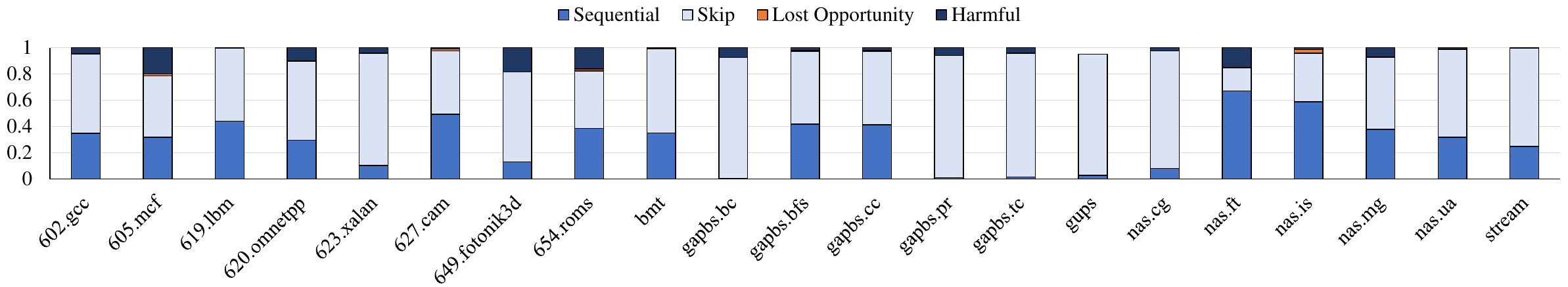}
  \caption{Breakdown of level prediction accuracy. Useful is the accesses that skip at least one level correctly. Opportunity loss is the fraction of lookups the could have avoided. Harmful is the fraction misprediction in all lookups.   }
\label{fig:F1}
\end{figure*}
\begin{figure}[t]
\centering
\includegraphics[width=\columnwidth]{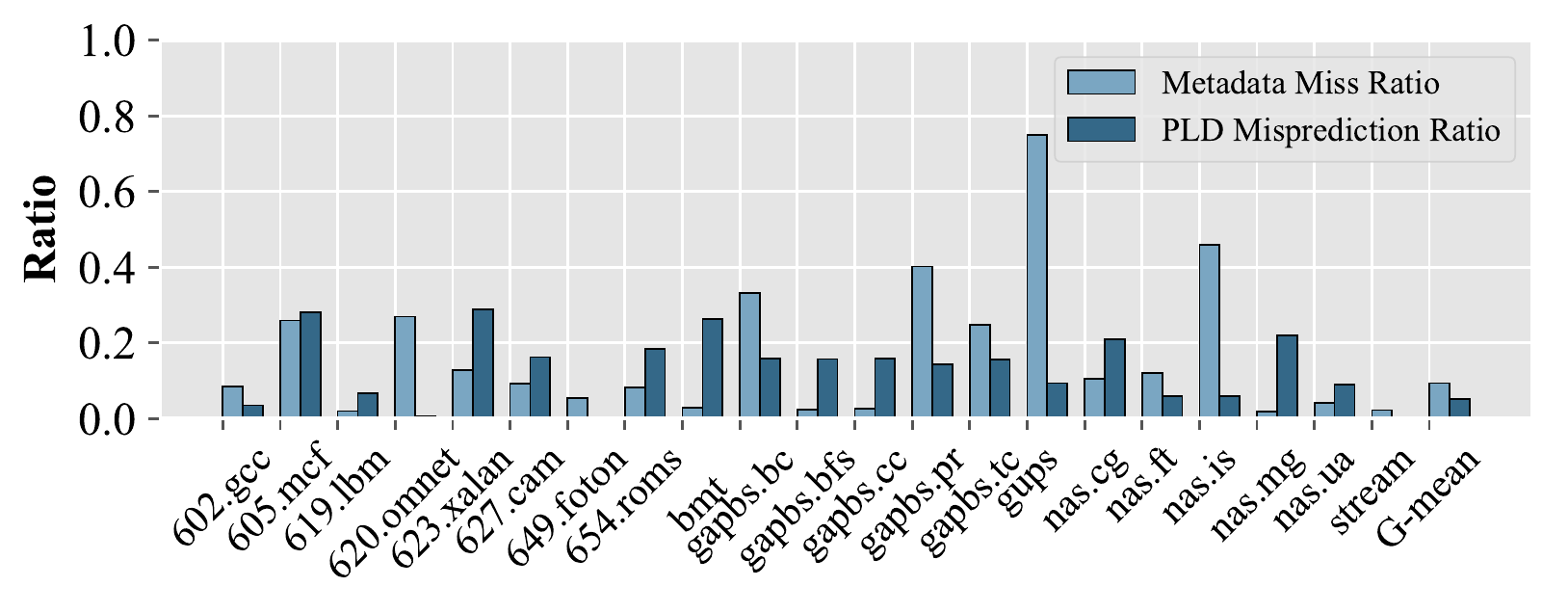}
  \caption{Metadata cache miss ratio and misprediction ratio of the Popular Level Detector.}
\label{fig:metaMiss}
\end{figure}


\subsection{Prediction Accuracy}

The LP predictions may be: (1) correctly predicted sequential (sequential); (2) correctly predicted skip (skip); (3) wrongly predicted sequential (opportunity loss); or wrongly predicted skip requiring recovery (harmful). Additionally, some predictions are multi-way and add some overhead despite reducing access latency. The overall prediction accuracy (\fig{fig:F1}) is very high. Only 605.mcf, 609.foto, 620.omnetpp, 654.roms, and nas.ft exhibit accuracies under 90\%, and only gups and and nas.is exhibit non-negligible lost opportunities. It is also clear that LP correctly identifies a large number of useful non-sequential lookup opportunities.

Interestingly, the lower accuracies are not a result of high \catalog metadata cache miss rates. As shown in \fig{fig:metaMiss}, those applications with lower accuracies still exhibit reasonable metadata cache hit ratios and high accuracy LDP predictions. Instead, the mispredictions are a result of stale \catalog information originating from a combination of aggressive prefetchers and poor metadata cache locality. When the prefetchers are aggressive, more clean lines are evicted without updating the \catalog. Furthermore, when the metadata cache replacement rate is high, a larger number of prefetch cache fills miss in the metadata cache and also do not update the \catalog.

\fig{fig:metaMiss} also demonstrates the importance of the Popular Levels Detector (PLD). Several benchmarks (605.mcf, 620.omnet, gapbs.bc, gapbs.pr, gups, and nas.is) exhibit high \catalog metadata cache miss rates, but those misses use the PLD, which offers high accuracy for these benchmarks. Note that the figure shows the PLD accuracy only considering those accesses that use it (metadata cache misses).

There are two reasons for the high accuracy of the PLD. The first is that some applications exhibit clear cache-level usage patterns, as discussed in \sect{sec:motivation} and summarized in \fig{fig:appsDomain}. For example, gapbs.bc and gapbs.tc exhibit very low hit rates in both L2 and L3, allowing the PLD to frequently suggest skipping both levels.

The second reason is that the PLD can suggest multi-way access to multiple levels in parallel, and thus not mispredict but with access overhead. This is shown in \fig{fig:codeStats} with good examples being: 620.omnet with 25\% of PLD predictions suggesting all levels in parallel, gapbs.pr with 35\% of PLD predictions of parallel accesses to memory and L3, and nas.is with 50\% of PLD predictions requesting simultaneous access to L2 and L3. However, for the most part multi-way prediction is rare.

\begin{figure}[t]
\centering
\includegraphics[width=\columnwidth]{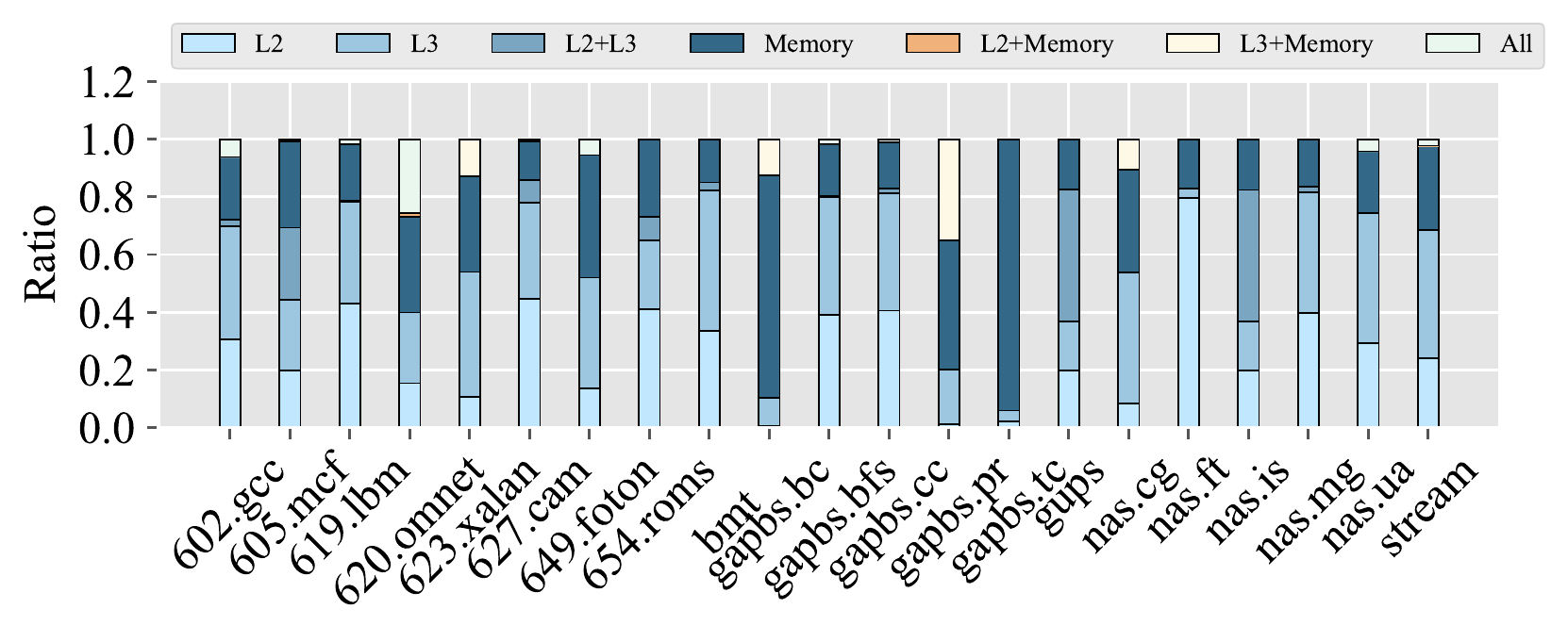}
  \caption{Levels suggested by the level predictor.}
\label{fig:codeStats}
\end{figure}

\begin{figure*}[t]
\centering
\includegraphics[width=\textwidth]{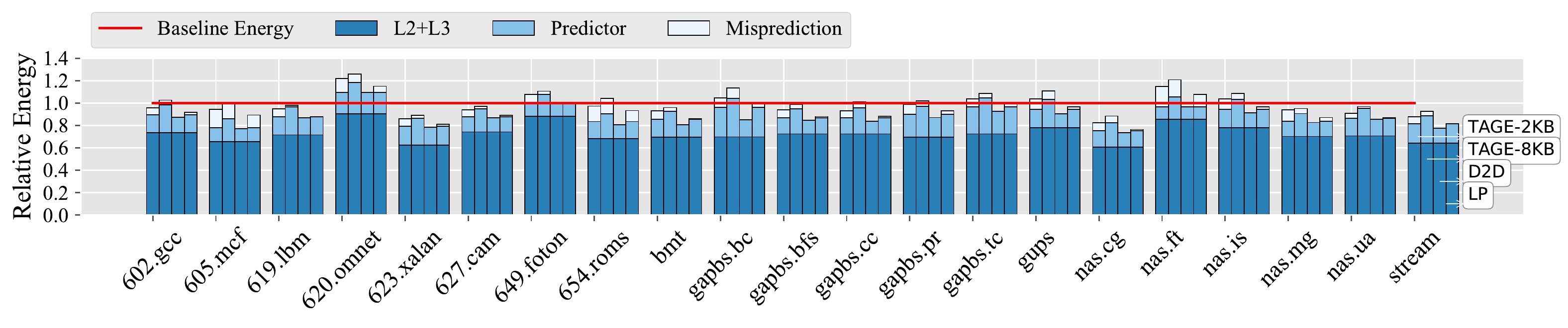}
  \caption{Energy consumption normalized to the baseline. For each benchmark, the bars are 8KB-TAGE, 2KB-TAGE, D2M, and level prediction from left to right. Ideal is "L2+L3" cache energy only.}
\label{fig:energy}
\end{figure*}




\subsection{Cache Energy Dissipation}
We use CACTI to obtain energy per access and then accumulate the total energy.
There are two major contributors to the energy consumption of predictors: (1) how frequent the structures are accessed to update and for prediction, and (2) how frequently we have to refer to the directory because of mispredictions. D2M also has two sources of energy overhead: (1) accessing larger TLB entries, and (2) updating the Hub on TLB misses and new insertions. While there are no mispredictions because D2M is a precise scheme, applications with high TLB miss rate (e.g. \emph{is}) need to access the hub more frequently, and thereby the energy consumption increases. Note that our energy analysis here refers to access energy alone and does not account for the additional energy savings resulting from the higher performance provided by level prediction.

\fig{fig:energy} shows the energy consumption of LP normalized to the baseline system and also compares this energy to the 2KB and 8KB address+history TAGE variants. We make four important observations. First, our \catalog + PLD predictor is substantially better than either TAGE variant. The 2KB TAGE has the same access energy as the LP, but its accuracy is far lower, which increases recovery overhead. In contrast, the 8KB TAGE offers similar (just slightly lower) prediction accuracy, but its access energy is far higher, resulting in significant additional cache-hierarchy energy.

The second observation is that our LP saves energy in all but two cases, resulting in an average energy saving of 16\%. The predictor is so accurate and the recovery scheme simple, such that on average only 1\% of the cache-hierarchy energy is spent on recovery.

The third observation is that in the two cases where energy is slightly increased, the overhead was a result of the very small benefit opportunity available. In 620.omnet, the energy overhead was the result of frequent all-level predictions by the PLD, while for nas.ft the overhead was a result of the low potential coupled with a relatively low prediction accuracy of just 08\%, and nearly all of those were for sequential lookup.

Finally, while not suffering from any mispredictions, D2M also exhibits overheads and can only improve on our LP by 3\% on average.



\begin{figure*}[t]
\centering
\includegraphics[width=\textwidth]{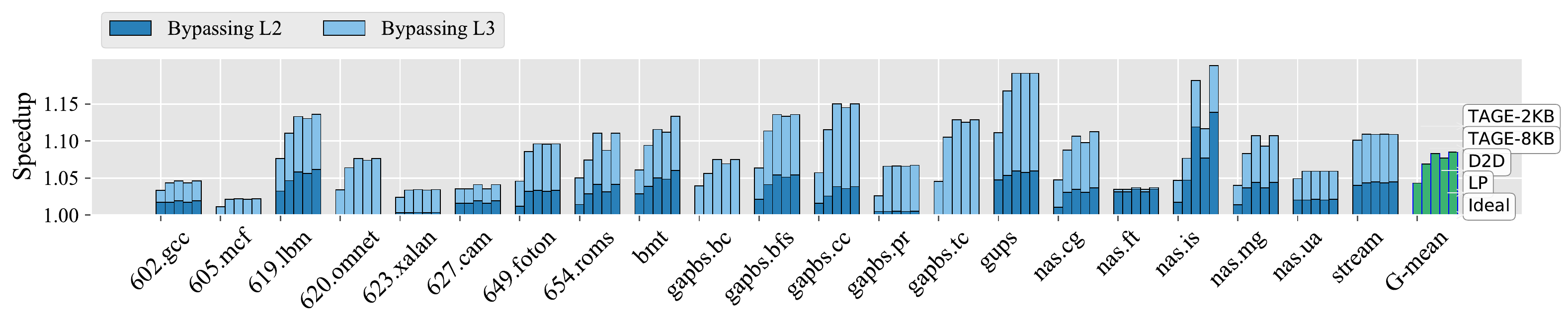}
  \caption{IPC improvement for a single core. For each benchmark, the bars are 2KB-TAGE, 8KB-TAGE, D2M, level prediction, and Ideal from left to right. }
\label{fig:IPCSignle}
\end{figure*}
\begin{figure}[t]
\centering
\includegraphics[width=\columnwidth]{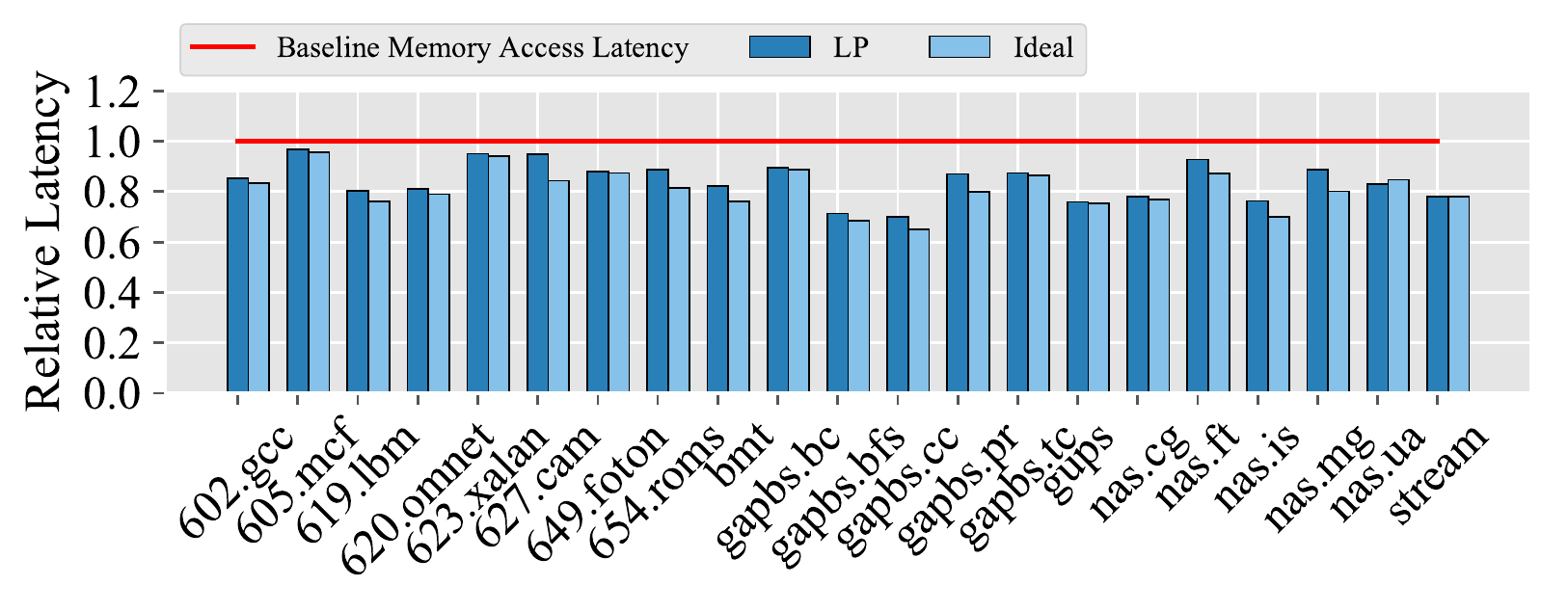}
  \caption{Memory access latency for single-core evaluation.}
\label{fig:MAL}
\end{figure}


\subsection{Performance}

\fig{fig:IPCSignle} shows the IPC improvement for the 2KB and 8KB TAGE predictors, D2M, our LP, and the idealized system. The geomean speedups are 4.3\%, 6.9\%, 8.2\%, 7.8\% and 8.4\%, respectively for each system. LP is within 10\% of the ideal speedup and trails the far more intrusive D2M architecture by just 5\% on average. At the same time, LP offers far better performance than building directly on the miss-predictor approach. We make four main observations on these results.

First, with two exceptions, speedup correlates well with the level-prediction potential discussed in \sect{sec:motivation} and summarized in \fig{fig:appsDomain}.
The largest speedup is achieved for those applications for which sequential lookup is harmful (application within the green box in \fig{fig:appsDomain}): 619.lbm, 649.foton, all the gpabs applications, and gups. The first exception is 605.mcf where speedup is just 3\%. The results of a top-down microarchitecture analysis~\cite{TAM} show that 619.lbm is bound by both memory bound (35\%) and the front-end ({31\%}). This, together with relatively high memory-level parallelism, limits the potential benefit of level prediction.
The second exception is nas.is, which falls outside the green box of \fig{fig:appsDomain} but still exhibits high speedup. We attribute this anomaly to the better prefetchers available on the commercial Intel core compared to our simulated processor. Indeed, when performing the same analysis with our simulated result, nas.is falls within the green box.


Our second important observation is that LP nearly matches the speedup of Ideal and D2M in all but two cases: 650.roms and nas.is. Both benchmarks exhibit high speedup potential and lower prediction accuracy compared to other applications with high-speedup potential. As shown in \fig{fig:F1}, LP only successfully skips a relatively small fraction of accesses (40\%) while also requiring recovery relatively frequently (20\%). The other two similar applications are 605.mcf and nas.ft, but those offer minimal speedup opportunity.
The reason for the large speedup difference with nas.is is different. For nas.is, the LP relies heavily on the PLD, which frequently suggests parallel L2 and L3 access. We argue that this increases pressure on the cache ports, realizing lower speedup than Ideal and D2D which do not attempt parallel accesses.

Our third observation is that while skipping L3 lookup and directly accessing memory instead has higher speedup benefits than skipping the much-lower latency L2, skipping L2 is still very useful. Just skipping L2 offers $>5\%$ speedup for many applications.



Finally, \fig{fig:MAL} shows the average memory access latency for LP and Ideal. The baseline is shown with a red line. The average memory access latency is improved by 20\% on average. Graph applications obtain lower memory access latency with level prediction because they have high miss ratios at all levels and avoiding those unnecessary lookups helps reduce  memory access latency. The trends match the speedup trends overall.

\begin{figure}[t]
\centering
\includegraphics[width=\columnwidth]{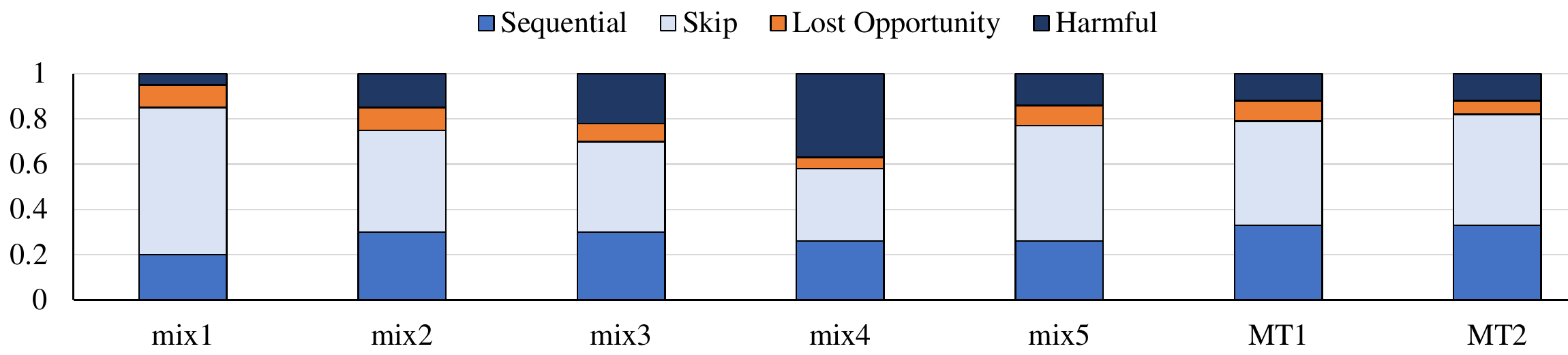}
  \caption{Multi-core level prediction accuracy.}
\label{fig:mpF1}
\end{figure}

\begin{figure}[t]
\centering
\includegraphics[width=\columnwidth]{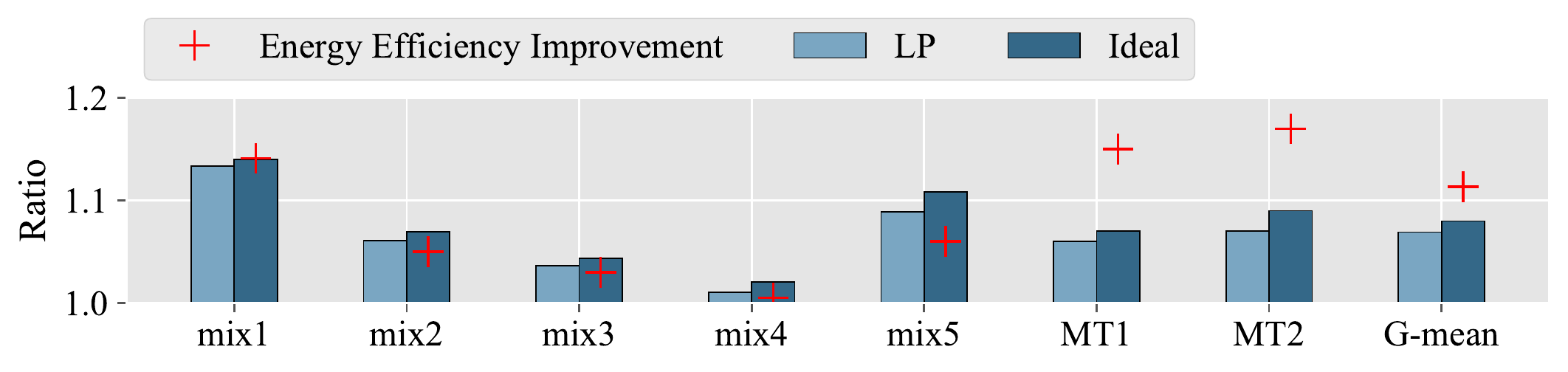}
  \caption{IPC and energy consumption of the multi-core system normalized to the baseline of Table {\ref{Table:System}}.}
\label{fig:mpIPC}
\end{figure}

\begin{figure}[t]
\centering
\includegraphics[width=2.7in]{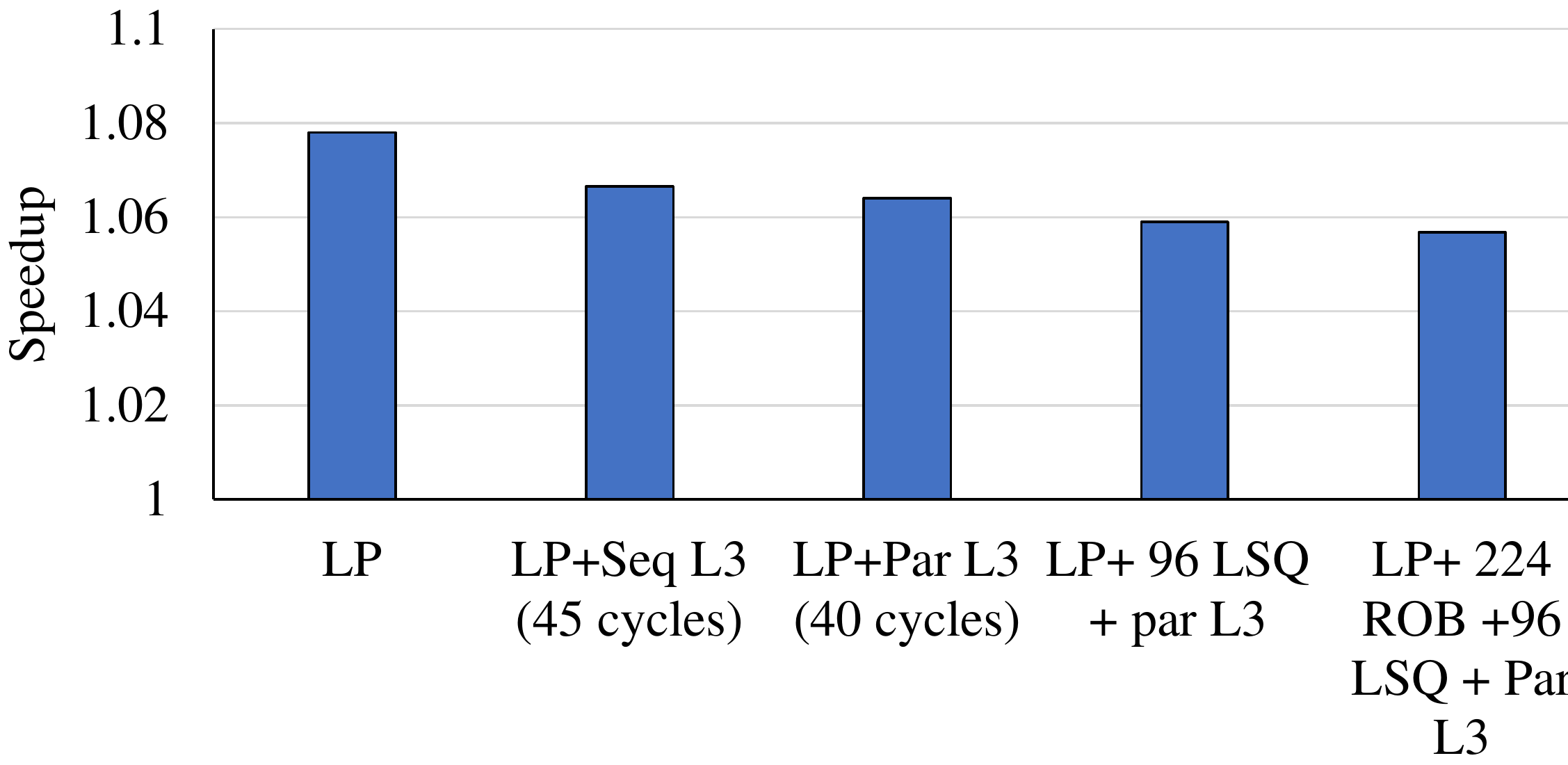}
  \caption{Sensitivity analysis to ROB size, LSQ size and LLC latency for a single core evaluation (average of all benchmarks).}
\label{fig:sens}
\end{figure}

\subsection{Multi-Core Results}
For multi-core simulation, we enable one LP per core. \fig{fig:mpF1} shows the predictor accuracy for the five multi-process mixes and the multi-threaded applications. Overall, LP accuracy with four cores is lower than with a single core. This is because contention on LLC is greater while prefetch aggressiveness is also substantially larger because more prefetchers operate in parallel. Still, accuracy is high with the exception of \emph{mix4}.  
For multi-threaded applications, we run gapbs.pr with 2 and 4 threads on a 4-core processor. While accuracy changes little between 2 and 4 threads, both harmful and opportunity-loss mispredictions are more frequent than with a single thread. This is expected because not only is there the greater LLC contention and prefetch aggressiveness, but there is also some degradation in \catalog accuracy because the \catalog is not updated with coherence events and because the \catalog uses a single entry per block even though it may or may not be cached in multiple private L2 caches.



\fig{fig:mpIPC} shows the speedup and relative energy-efficiency improvement for the mixes. Notably, level prediction always provides some speedup and cache-energy improvement. For multi-program mixes, \emph{mix1} has the highest IPC improvement (13\%), and \emph{mix4} has the lowest (1.2\%). The main reason is that \emph{mix4} is composed of four low-MPKI applications and offers minimal speedup potential, whereas \emph{mix1} has 4 very high MPKI applications and offers high speedup potential. Overall, the speedup geomean with LP is 6\%, achieving a large fraction of the potential 7\% geomean speedup of the idealized system. Energy efficiency is improved by an average (geomean) of 8\%, which is again, more than 85\% of the potential energy benefits of the ideal case. 

For multi-threaded applications, speedup improves as the thread count increases. This is despite the prediction accuracy slightly decreasing because of higher LLC contention. This same LLC contention, however, also increases speedup potential because memory is accessed mmore frequently.  

\subsection{Sensitivity Analysis}
Figure {\ref{fig:sens}} compares the average normalized IPC across all applications for 5 systems: (1) a system with the configuration of Section~\ref{sec:eval}, (2) a system with fast sequential LLC (45 cycles), (3) a system with parallel LLC (40 cycles of access latency), (4) a system with a parallel LLC, and a larger LSQ (96 entries), and (5) a system with a very aggressive core (ROB=224, and LSQ=96) and a parallel LLC. To be fair, for each scenario, we normalize the IPC to the baseline with that configuration without level prediction. Speedup geomeans of the 5 systems are 7.8\%, 6.6\%, 6.4\%, 5.9\%, and 5.6\%. As can be seen, the overall improvement decreases as the systems become more aggressive. However, even for the most aggressive configuration, level prediction provides 5.6\% speedup compared to the clearly aggressive baseline. 

\subsection{Overhead Analysis}
Our design requires only a 2KiB metadata cache per core as well as three 32-bit wide counters. For each 64B-block in physical memory 2 bits are assigned leading to a memory overhead of 0.39\%. The directory remains unchanged, and only the cache controller is notified with a mechanism to deallocate the MSHR entries on a misprediction.


\section{Related Work}
\label{sec:relatedwork}

Per level hit/miss predictor is either used in front of L1 cache \cite{MattanHitMiss, PeirHitMiss} to handle 
instruction scheduling better, or only employed at L4 caches \cite{MissMap, MoinL4, Mostly-Clean}. The insight behind L4 hit/miss prediction that the miss penalty is high, and blindly accessing cache incurs high performance and power consumption costs. Our proposal extends such a solution to all memory hierarchy, which is getting deeper recently.

D2M \cite{SplitMeta} finds the actual location of a block in the memory hierarchy with a single lookup. Authors separate the metadata from the data hierarchy, and then using pointers, the request is forwarded to its destination. This technique requires significant changes to the current processor, such as enlarging TLB entries and adopting a new coherence scheme. D2D is another technique to provide a 
single lookup in cache hierarchy \cite{D2D}, however, it also needs to enlarge the TLB entries and OS changes. 

Cache bypassing technique \cite{RTBypassing, GPUBypassing} selectively inserts data blocks in the cache. Because many applications with  streaming behavior have no data resue, and they better to be written directly to the main memory, and keep the valuable on-chip space for incoming requests. Way prediction reduces energy consumption by avoiding searching all ways to match the tag 
\cite{WayPrediction}.

Software prefetching is an appealing solution to reduce memory access latency for applications with complex address patterns \cite{Sam_IMP, Sam_event, SAM_SWP}. 
This technique is useful when the memory access pattern is complex and thereby cannot be captured by hardware prefetchers 
\cite{Sam_IMP}. Many previous works attempted to handle indirect memory access patterns by inserting the software prefetch instruction in the code. Also, properly adjusting the prefetch distance is a crucial for a timely prefetch. 
Pro-actively finding the best time slot to issue the software prefetch request has been studied in \cite{SAM_grpah}. Event-driven software prefe tch generation has been proposed in \cite{Sam_event}. Although software prefetch can increase the coverage to almost 100\%, it 
severely suffers from lack of timeliness. 
its effectiveness.

Many prefetchers have been proposed in the past decade \cite{Sam_IMP, Sam_event, SAM_SWP}\cite{BOP,STeMS,ISB,DCPT,SBOOE,IMP,SPPV2,PIF,SlimAMPM,AMPM}.
Spatial prefetchers rely on spatial behavior of access patterns to predict the next address. However, temporal prefetchers 
record the past addresses and use them for future predictions. Spatial prefetchers require complex logics to detect the address 
patterns, while temporal prefetchers need a significant amount of metadata to record the past addresses. 
achieve at most 40\% of coverage \cite{Bingo}.


Cache level prediction can achieve high accuracy with a simple table. Also, in contrast to D2M \cite{SplitMeta}, the overall 
system design can remain untouched because it utilizes the available resources and schemes. Additionally, compared to a sequential per level predictors where separate 
predictors for each level can sit close to CPU, our unified predictor takes smaller space and attain better accuracy as it 
aggregates information all in one table, and observes the system as whole.

\section{Conclusion}
\label{sec:conclusion}
We propose a cache level predictor in order to enhance the lookup  strategy  in  multiple-cache setting. This technique filters unnecessary accesses to intermediate levels, and thus reduces the cumulative miss latency. The proposed system enhances the IPC by 7.8\% compared to the baseline.

\bibliographystyle{IEEEtranS}
\bibliography{references}
\end{document}